\documentclass[sigconf]{acmart}

\setcopyright{cc}
\setcctype{by}
\copyrightyear{2026}
\acmYear{2026}
\acmDOI{10.1145/3858063.3858747}
\acmConference[ICAIF '26]{7th ACM International Conference on AI in Finance}{November 14--17, 2026}{Milan, AA, Italy}
\acmBooktitle{7th ACM International Conference on AI in Finance (ICAIF '26), November 14--17, 2026, Milan, AA, Italy}
\acmISBN{979-8-4007-3087-0/2026/11}

\begin{document}

\title{Spreads Don't Price a Rival Market Maker's Information}

\author{Oliver James Oberle}
\correspondingauthor
\orcid{0009-0003-6920-3455}
\affiliation{
  \institution{ETH Z\"urich}
  \city{Z\"urich}
  \country{Switzerland}}
\email{ooberle@ethz.ch}

\author{Damien Berriaud}
\correspondingauthor
\orcid{0009-0002-0450-0339}
\affiliation{
  \institution{ETH Z\"urich}
  \city{Z\"urich}
  \country{Switzerland}}
\email{dberriaud@ethz.ch}

\author{Roger Wattenhofer}
\orcid{0000-0002-6339-3134}
\affiliation{
  \institution{ETH Z\"urich}
  \city{Z\"urich}
  \country{Switzerland}}
\email{wattenhofer@ethz.ch}

\begin{abstract}
When one market maker becomes better informed, do its competitors widen their bid-ask spreads? Adverse-selection reasoning suggests that they should, because the better-informed maker avoids toxic order flow and leaves more of it to the others, and the closest model of competition with an informed dealer predicts more conservative quotes by the uninformed dealers. We study this question in a stochastic quoting game in which one maker receives a private signal of varying precision about a persistent state that drives order flow. Strategies are learned with independent reinforcement learning, which we benchmark against an exact dynamic-programming solution of the single-agent case. In none of the configurations we examine does the average spread of the uninformed makers increase with the rival's precision. Where competition for order flow is weak, the slope of their spread in precision is statistically equivalent to zero within half an action tick; where competition is strong, their spread tightens. The informed maker's own spread tightens markedly in every configuration and serves as a positive control. These results hold with two, three and four makers, across an eightfold variation in the depth of the background liquidity, with a background quote that reprices after toxic fills, and under price-time priority. Our measurements point to a competitive trade-off. In the model the total toxicity of the flow does not depend on any maker's information. A better-informed maker avoids more of it, and where this leaves more toxic flow to its competitors, it also quotes more tightly, so that a wider quote would cost them more order flow than it would save in adverse selection. The uninformed makers adjust the skew of their quotes and their inventory instead, and the rival's information changes the distribution of profits across makers while quoted spreads remain nearly unchanged.
\end{abstract}


\maketitle

\section{Introduction}\label{sec:intro}
Market makers supply liquidity. Each maker posts a price at which it will buy and a price at which it will sell, on either side of a fluctuating market price, and competes with other makers for the traders who take these quotes. A maker earns the distance between its quote and the market price on every unit it trades and loses when the price subsequently moves against the resulting position. Order flow that tends to arrive just before such a move is called \emph{toxic}, and it is the main reason why quotes cannot be arbitrarily tight.

When one maker becomes better informed about the direction of the next price move, two effects act on the spreads of its competitors. We call the first the \emph{protective} force. The better-informed maker trades more of the harmless flow and less of the toxic flow, so the flow left to the other makers is worse on average. The adverse-selection logic of models with informed takers \citep{glosten1985, kyle1985} then implies that the others should widen their quotes to break even. We call the second the \emph{competitive} force. The better-informed maker quotes more tightly and wins more fills, which pushes the others to tighten in order to keep their share of the flow. Which effect dominates is an open question. Informed-taker theory does not carry over directly, because an informed \emph{competitor}, unlike an informed customer, never trades against the other makers' quotes. Private signals, inventories and the public history of quotes all enter each maker's strategy, and the resulting repeated stochastic game has no closed-form solution.

We therefore study the question numerically. We formulate a reduced-form quoting game in which one maker holds a private signal about a persistent latent state that drives order flow, all quotes are public, and execution is a stochastic contest for a common stream of external market orders. Non-strategic background liquidity, which we call the \emph{fringe}, takes part in the same contest. Multi-agent reinforcement learning (MARL) approximates stationary strategies, and we compare outcomes across values of a single parameter, the precision of the informed maker's signal. We say that a spread \emph{prices} a quantity if it responds to it, so our question concerns the sign of the slope of the uninformed spread in the rival's precision.

In every configuration we study, this slope is either statistically equivalent to zero or negative. When competing for the marginal fill is unattractive, either because the fringe quotes so tightly that undercutting it attracts mostly toxic flow or because it quotes so widely that the makers' rents are safe, the uninformed spread is flat within a margin fixed in advance. When the marginal fill is contested, because the fringe is thin or quotes close to the strategic makers, the uninformed makers tighten as the rival's quotes tighten. This is a response to the rival's observable quotes. We find no configuration in which the uninformed spread widens with the rival's precision. The rival's precision does, however, change who earns the rents and who bears the adverse selection.

Three features of the model limit the scope of these results. First, external market orders arrive as a price-inelastic Poisson stream whose toxicity does not depend on the informed maker's precision, so information asymmetry exists only among the makers. Second, execution is a contest with exponential weights, in which undercutting raises a maker's share of the flow smoothly and never captures all of it; Section~\ref{sec:ptp} retrains the game under price-time priority to check whether this choice drives the result. Third, the fringe absorbs part of any flow that the strategic makers decline; Section~\ref{sec:af} replaces it with an \emph{adaptive} fringe that reprices according to its own realised toxicity. Our results therefore concern oligopolistic markets with a competitive fringe.

The closest model of price competition with an informed dealer points toward widening: in \citet{CalcagnoLovo2006}, uninformed dealers quote more conservatively because they face a winner's curse against the informed dealer. None of the models discussed in Section~\ref{sec:related} varies the precision of a rival's information. We contribute a measurement of the response, evidence on the mechanism behind it, and an account of the features of our setting that offset the widening.
(1)~\emph{An equivalence test with a positive control.} The average spread of the uninformed makers does not increase with the rival's precision in any configuration. Where the fringe leaves little to compete for, the per-seed regression slope of spread on precision is equivalent to zero within half an action tick under two one-sided tests (TOST). 
Every configuration that does not pass the test has a negative point estimate; only one of them, the thinnest fringe, is significantly negative.
In the same sweeps the informed maker's spread tightens strongly and fails the equivalence test, so the design does detect a response where one is present.
(2)~\emph{Evidence on the mechanism across competitive regimes.} A quoted spread reflects both the toxicity a maker expects on its fills and how strongly it competes for them, and a better-informed rival affects these two considerations in opposite directions. It worsens the flow left to the others, which favours wider quotes, and it quotes more tightly, which increases the flow lost per unit of additional width. In our experiments the second effect dominates. Under the baseline the first effect is close to zero, because the fringe absorbs the toxic flow that the informed maker avoids. With a repricing fringe the uninformed makers do receive more toxic flow, and the second effect still dominates. At $p_1 = 1.0$, a constant quote tuned to the low-precision regime falls $24.2$ payoff units short of the learned uninformed policy; adding a defensive-widening term reduces this shortfall only to $18.2$, whereas adding a term that follows the rival's direction reduces it to $4.5$. The value of the signal is thus reflected in the distribution of profits, and quoted spreads change little.
(3)~\emph{A validated and audited learning pipeline.} The single-agent version of the game has an exact dynamic-programming solution, which the reinforcement learner reproduces closely before we make any multi-agent claim. Because learned profiles need not be equilibria \citep{ContXiong2024}, we audit the profiles behind the main claims for exploitability and report the results where they qualify a claim.

\section{Related Work}\label{sec:related}
A market maker's spread compensates it for two risks: holding inventory while the price moves \citep{HoStoll1981, AvellanedaStoikov2008}, and trading with better-informed counterparties \citep{glosten1985, kyle1985}. In the second line of work the informed party is a \emph{taker}, a customer who trades against the maker's quote. The competitive spread is then the break-even compensation for losses to that flow, and it widens as informed trading increases. \citet{BarucciMathieuSB2025} bring informed takers into the inventory-based framework of \citet{AvellanedaStoikov2008}. A single market maker trades with informed and uninformed traders; the arrival rates of the informed traders depend on a fad, a mean-reverting deviation of the mid-price from fundamental value, and the maker cannot tell the two groups apart. Holding the expected number of arrivals fixed, they find that the maker's performance falls as the share of informed traders grows, while its spread increases slightly, and that information about the fad is valuable to the maker; in their results, this information shows up mainly in the skew of the quotes and in the maker's performance. Their fad plays a role similar to our state $z_t$, which also tilts the arrival rates; it enters their mid-price directly, whereas $z_t$ moves our mid only through the realised order imbalance, so that in both models knowledge of the latent state predicts the next price move. In informed-taker models such as theirs and that of \citet{glosten1985}, the takers' information and the informed share determine how much toxicity the flow carries. In our setting the informed party is a competing \emph{maker}, which does not trade against its rivals' quotes, and the toxicity of the flow does not depend on any maker's information. The two settings are thus complementary: theirs varies how toxic the flow is and how much a single maker knows about it, ours varies which of several makers bears a fixed amount of toxicity.

The theory closest to our question concerns price competition between asymmetrically informed dealers. \citet{Dennert1993} shows that, when informed insiders trade, adding market makers raises each maker's risk exposure and can widen individual spreads; this channel runs through informed takers, which our model excludes. \citet{BiaisMartimortRochet2000} characterise competition in supply schedules, where rents shrink with the number of competitors but remain positive for any finite number. \citet{CalcagnoLovo2006} study repeated bid-ask competition between one dealer that knows the asset value and uninformed dealers, with uninformed takers and price priority. The informed dealer's identity is common knowledge, and the uninformed dealers learn from its public quotes, so that their quotes follow its earlier quotes (price leadership); the informed dealer earns positive profits and the uninformed dealers earn zero. Because a quote that wins against a better-informed dealer tends to be the mispriced one, the uninformed dealers face a winner's curse and quote more conservatively. The spread is therefore positive although no taker is informed, whereas it would be zero under symmetric information, and in their numerical example the expected inside spread widens as the public announcement approaches and the informed dealer's quotes become more informative. Read as a comparative static in the informativeness of the rival's quotes, their mechanism points to the defensive widening that we do not find. Our setting differs in several respects that bear on this mechanism. Execution is a contest that never awards the whole flow to the best quote, whereas theirs is a first-price auction; a fringe absorbs flow that the strategic makers decline; the uninformed makers keep positive rents; our makers carry inventory, which their risk-neutral dealers do not; and our signal concerns a Markov state and is drawn afresh at every step, whereas their informed dealer holds long-lived information until a public announcement and may deliberately mislead the others. Our price-time priority treatment (Section~\ref{sec:ptp}) removes the first of these differences, and the result persists. Their informed dealer is perfectly informed, and their model has no comparative static in the precision of its information, the parameter we vary. We agree with them on price leadership and on positive rents for the informed side. Our treatment with public role labels relates to the study of anonymity in limit order books by \citet{FoucaultMoinasTheissen2007}.

Reinforcement-learning market makers facing strategic competitors were introduced by \citet{Ganesh2019}, and multi-agent learning has since been developed as a simulation methodology for over-the-counter markets \citep{Vadori2024}. \citet{Bernasconi2023} compute approximate equilibria of a dealer market with many strategic dealers by combining reinforcement learning with mean-field games, and test the learned strategies against dealers trained to exploit them. The mean-field limit treats each dealer as negligible; our setting is a finite oligopoly in which the information of one identifiable rival varies. Their exploitation test is close in spirit to our best-response audit. \citet{ContXiong2024} show that independent learners in a dealer game can settle at spreads above the competitive level, which illustrates that learned profiles need not be equilibria. \citet{Jerome2023} build their environment around validating learners against exact benchmarks, an approach we follow. For verification we combine the deviation-payoff estimation of empirical game-theoretic analysis \citep{WahWrightWellman2017}, exploitability as a measure of equilibrium quality \citep{Lanctot2017}, and best-response regret as in \citet{Dwarakanath2024}. We treat the learned joint policy as a hypothesis about approximate equilibrium play and test it. To our knowledge, neither the theoretical nor the learning literature varies the quality of a rival maker's information; it is either absent, fixed at perfect, or proxied by the number of competitors.

\section{Model}\label{sec:model}
The environment is a stylised limit order book run for $T$ discrete steps. Each step proceeds in the same order: the mid-price $m_t$ is publicly observed; the informed maker receives its private signal; all makers simultaneously post one bid and one ask; the step's market orders arrive and are allocated to quotes; inventories are updated; and the mid moves to $m_{t+1}$. There are $N \in \{2,3,4\}$ strategic makers, indexed by $i$. Maker $i$ chooses two non-negative distances $(\delta^b_{i,t}, \delta^a_{i,t})$, offering to buy at $m_t - \delta^b_{i,t}$ and to sell at $m_t + \delta^a_{i,t}$, with capacity $u$ units per side. External traders, called \emph{takers} because they consume posted liquidity, arrive at random and trade against these quotes. All quotes are public; signals and inventories are private.

\paragraph{Order flow and prices.}
A persistent latent pressure state $z_t \in \{-1, +1\}$ evolves as a two-state Markov chain with $\Pr(z_{t+1} = z_t) = \rho \in (\tfrac12, 1)$. Conditional on $z_t$, external buy and sell market orders arrive as $M_t^{\pm} \sim \mathrm{Poisson}(\Lambda_t^{\pm})$ with $\Lambda_t^{\pm} = \bar\Lambda e^{\pm\gamma z_t}$, so that buy orders outnumber sell orders on average when $z_t = +1$, and the reverse holds when $z_t = -1$. The mid-price moves with the realised signed imbalance plus noise,
\begin{equation}
m_{t+1} = m_t + \phi\,(M_t^+ - M_t^-) + \sigma\,\varepsilon_{t+1}, \qquad \varepsilon_{t+1} \sim \mathcal N(0,1),
\end{equation}
with constants $\phi, \sigma > 0$, so $z_t$ induces a predictable short-run drift $\mathbb{E}[m_{t+1} - m_t \mid z_t] = \phi(\Lambda_t^+ - \Lambda_t^-)$, about $\pm 0.78$ per step at the baseline parameters of Section~\ref{sec:method}. Knowing $z_t$ therefore means knowing the likely direction of the next price move. This drift makes one side of the flow \emph{toxic}. At $z_t = +1$ the abundant buy orders lift ask quotes, so a maker filled on its ask has sold just before the price rises, while fills on its bid are harmless. Which side is toxic depends only on $z_t$. No parameter of any maker's information enters the flow process, so the total toxicity in the market is fixed.

\paragraph{Information.}
Maker $1$ is informed. At each step it receives a private signal $x_{1,t} \in \{-1,+1\}$ with $\Pr(x_{1,t} = z_t) = p_1 \in [\tfrac12, 1]$. The remaining makers receive no signal, and we set $x_{i,t} = 0$ for them. We call $p_1$ the \emph{precision} of the signal; it ranges from an uninformative signal ($p_1 = \tfrac12$) to perfect information ($p_1 = 1$). Only one maker is informed in each episode, and the others can learn about its signal only through its quotes, which makes quote placement strategic. The informed role is fixed in most treatments and reassigned at random in each episode in one (Section~\ref{sec:flat}). We measure directional quoting by the quote asymmetry, or \emph{skew}, $s_{i,t} = \delta^b_{i,t} - \delta^a_{i,t}$; a maker that tightens one side and widens the other has a non-zero skew. Under this convention a maker that tightens its bid ahead of expected buy pressure ($z = +1$) has $\mathrm{corr}(s, z) < 0$.

\paragraph{Execution.}
Let $\mathcal{A}_t$ be the set of ask quotes still available at step $t$, consisting of the $N$ makers' asks and the fringe's ask, each with exponential execution weight $w_j = e^{-k\delta_j}$. Each of the $M^+_t$ buy orders is assigned to one of these quotes with probability
\begin{equation}
\Pr(\text{order} \rightarrow j) \;=\; w_j \Big/ \textstyle\sum_{l \in \mathcal{A}_t} w_l ,
\end{equation}
after which the remaining capacity of that quote falls by one unit; a quote leaves $\mathcal{A}_t$ once its capacity is exhausted. Sell orders are allocated to bid quotes in the same way. Writing $F^a_{i,t}$ and $F^b_{i,t}$ for the number of units maker $i$ trades on its ask and on its bid, inventory evolves as $q_{i,t+1} = q_{i,t} + F^b_{i,t} - F^a_{i,t}$. It is bounded by $|q| \le q_{\max}$: a maker at its long limit stops bidding until it has sold, and symmetrically at its short limit. A quote closer to the mid is exponentially more likely to be filled, but no quote is filled with certainty, so tightening raises a maker's expected share of the flow smoothly.

We use this stochastic allocation deliberately. Each discrete period aggregates a race within the period that we do not model, and the lottery approximates its outcome in real markets, where latency and queue position often allow a slightly worse price posted earlier to trade first. The allocation removes the winner-takes-all margin at the touch, and with it the winner's-curse channel of markets with strict price priority. It preserves the two properties our question requires: relative aggressiveness shifts fill shares, and quote placement therefore carries information. Comparable formulations appear in \citet{BoyceHerdegenSB2025}, and Section~\ref{sec:ptp} retrains the game under price-time priority.

\paragraph{The fringe.}
A non-strategic background quote on each side, the \emph{fringe}, takes part in the same allocation with weight $e^{-k\delta_{\mathrm{bg}}}$. It represents the remaining resting liquidity in the market, supplied by participants who follow simple rules. The fringe is uninformed and quotes symmetrically at $\delta_{\mathrm{bg}} = 1/k$, the distance that maximises the expected revenue per step $\delta e^{-k\delta}$ and hence the optimal symmetric quote of the single-dealer model in its risk-neutral limit \citep{AvellanedaStoikov2008}. It does not condition on $z_t$, holds no inventory and, in the baseline, does not move. The fringe has two roles. It sets a floor on competition, because the strategic makers always compete against a quote at this distance. It also gives takers an alternative, so flow that the strategic makers decline is absorbed by the fringe instead of being lost. Sections~\ref{sec:structure} and~\ref{sec:af} vary both roles.

\paragraph{Reward.}
Each maker's per-period reward takes the standard running form of the optimal market-making literature \citep{AvellanedaStoikov2008, CarteaJaimungalPenalva2015},
\begin{equation}
r_{i,t} = F^a_{i,t}\,\delta^a_{i,t} + F^b_{i,t}\,\delta^b_{i,t} + q_{i,t}\,(m_{t+1} - m_t) - \eta\, q_{i,t}^2 .
\end{equation}
Each filled unit earns its quoted distance (spread capture), the inventory gains or loses as the mid moves (mark-to-market), and the quadratic term represents risk aversion. With $\eta = 0$ and an episode that ends flat ($q_T = 0$), the first three terms sum to the realised cash profit, so the running form is a decomposition of that profit and does not define a separate objective. Agents maximise the undiscounted episode return. Information about $z_t$ earns money only through the mark-to-market term, and toxicity costs money through the same term: an adverse fill is one whose \emph{markout}, the price move that follows it, reduces this term. The term is charged on the inventory held before the step's fills, so a unit traded at $t$ is first marked at $t+1$. Fills arrive throughout a period, and charging them the full concurrent move would overstate their exposure as much as charging nothing would understate it. Because the state persists ($\mathbb{E}[z_{t+1} \mid z_t] = (2\rho - 1)\,z_t$), each position is still charged a fraction of about $2\rho - 1$ of its expected markout, $60\%$ at $\rho = 0.8$. This fraction is the same at every precision, so the convention cannot create or mask a response to $p_1$.

\section{Learning and Evaluation}\label{sec:method}
Baseline parameters throughout are $T = 200$, $q_{\max} = 10$, $\rho = 0.8$, $\bar\Lambda = 1.5$, $\gamma = 0.5$, $k = 1.5$, $\phi = 0.5$, $\sigma = 0.5$, $\eta = 0.01$, capacity $u = 1$ per side (raised to $u \in \{3, 5\}$ in Section~\ref{sec:structure}), and a $10 \times 10$ action grid over $[0.1, 2.0]$ per side. We refer to one configuration at one precision as a \emph{cell}; each cell is trained from ten independently seeded runs unless stated otherwise. Every configuration has exactly one informed maker, and every uninformed quantity we report is averaged over the $N-1$ uninformed makers.

\subsection{Exact Single-Agent Benchmark}\label{sec:anchor}
With one strategic maker facing the fringe, the state space $(t, q, z)$ is finite, and backward induction yields the optimal policy exactly. The omniscient maker ($p = 1$) earns $V(0) = 161.54$ per episode, compared with $117.91$ for an uninformed maker ($p = \tfrac12$), a value of information of $43.63$ ($+37.0\%$) that is monotone and concave in $p$. At $z = +1$ the optimal informed policy tightens the bid and widens the ask, giving a simulated $\mathrm{corr}(s, z) = -0.93$. A single-agent PPO learner \citep{Schulman2017}, trained on episodes that start from randomly drawn inventories so that every state is visited \citep{SuttonBarto2018}, recovers $94$ to $96\%$ of the exact optimum and $88$ to $94\%$ of the value of information across compute budgets. It also reproduces, state by state, the signal-driven asymmetry and the inventory mean reversion of \citet{HoStoll1981}. This checks the environment and the learning pipeline before any multi-agent result.

\subsection{Independent PPO}\label{sec:ippo}
Multi-agent policies are trained with \emph{independent} PPO \citep{Schulman2017, deWitt2020}. Each maker has its own actor and critic over its own observations and reward, with no shared critic and no information sharing during training. This matches the assumption that the makers are separate, competing firms that each optimise their own book; a shared critic or a team reward would introduce coordination that a competitive market lacks. Each agent observes its own $(q_{i,t}, x_{i,t}, T - t)$ together with a summary of each competitor's public quotes: the most recent quote distances, their rolling means over $W = 10$ steps, and a rolling correctness score $h_{j,t} = \tfrac{1}{W} \sum_{\ell} \mathrm{sign}(s_{j,\ell})\,\mathrm{sign}(m_{\ell+1} - m_\ell)$, which records whether competitor $j$'s skew has been aligned with realised price moves. Under our sign convention this score is negative for an informed maker. A quoting desk could compute all of these features from a public order-book feed. Networks are two-layer MLPs with 64 units per layer and hyperparameters taken from the validated single-agent configuration, and each run trains for $10^7$ environment steps. Policies are evaluated by Monte Carlo simulation from $q_0 = 0$ with $1{,}000$ episodes per seed, using a common stream of episode seeds across compared configurations so that all reported differences are paired.

\subsection{Audits}\label{sec:audit}
Comparative statics of learned play are informative only if the learned profile is close to an equilibrium, so we test this in four ways.
\emph{(i) Best-response regret.} We bound what a single maker could gain by deviating. We freeze the other makers, retrain the maker in question for $10^6$ further steps starting from its own weights, and compute the paired payoff difference over a common stream of episode seeds. The empirical noise floor is roughly 8 to 15 percent of an agent's payoff, and tripling the retraining budget changes the estimates by less than a fifth of a joint confidence interval, so the estimates do not depend on the budget.
\emph{(ii) Parametric deviations.} Deviation families suggested by theory, namely constant spreads and belief-conditioned rules, are evaluated against the frozen learned opponents by paired simulation, without training.
\emph{(iii) Equivalence testing.} A flat spread is a null claim. We therefore formulate it as equivalence and test it with two one-sided tests (TOST), which reject the hypothesis that the response lies outside a margin fixed in advance. Passing the test shows that the response is smaller than the margin; it does not show that the response is exactly zero. We require both one-sided tests to reject at the five percent level and report the larger of the two $p$-values. The margin is half an action tick, that is, half of the smallest change in spread an agent can make, which gives $0.105$ in spread units. The margin is thus set by the environment and was not calibrated on the results; it amounts to about four percent of the quoted spread. Where a configuration was trained at three or more precisions, we test the per-seed OLS slope of spread on $p_1$ against $\pm 0.222$ per unit of $p_1$. This corresponds to a change of $0.10$ over the swept range $[0.55, 1.0]$ and is slightly stricter than half a tick ($0.105/0.45 = 0.233$). Where only the two extremes were trained, we test the per-seed difference $\mathrm{spread}(p_1{=}1.0) - \mathrm{spread}(p_1{=}0.55)$ against $\pm 0.105$. In both cases the informed maker's spread serves as a positive control and should fail the same test.
\emph{(iv) Signal-blind rollouts.} Running a trained policy with its own signal input set to zero gives a direct estimate of what the signal is worth to that policy and identifies policies that would earn more by ignoring their signal.

\section{Results}\label{sec:results}

\subsection{Response of the Uninformed Spread}\label{sec:flat}
Table~\ref{tab:sweep} reports the three-agent game with fixed and commonly known roles: maker 1 is informed with precision $p_1$, and two makers are uninformed. We sweep $p_1 \in \{0.55, 0.7, 0.85, 1.0\}$ with ten seeds per precision and $1{,}000$ paired evaluation episodes per seed.

\begin{table}[t]
\centering
\caption{Fixed-role three-agent sweep across the informed maker's precision $p_1$. Ten independently trained seeds per precision, each evaluated over $1{,}000$ Monte Carlo episodes; entries are means $\pm$ the standard error of the mean (SEM) across seeds. Uninformed entries average the two uninformed makers. VoS is the informed-minus-uninformed payoff difference; spreads are the sum of the two per-side quote distances.}
\label{tab:sweep}
\footnotesize
\setlength{\tabcolsep}{4pt}
\begin{tabular}{lcccc}
\toprule
 & $p_1{=}0.55$ & $p_1{=}0.7$ & $p_1{=}0.85$ & $p_1{=}1.0$ \\
\midrule
Informed payoff & $46.9 \pm 1.4$ & $51.6 \pm 2.0$ & $66.0 \pm 2.0$ & $94.5 \pm 0.6$ \\
Uninf.\ payoff & $45.3 \pm 0.9$ & $45.5 \pm 1.3$ & $45.9 \pm 1.4$ & $39.3 \pm 1.3$ \\
VoS & $1.6 \pm 1.7$ & $6.2 \pm 1.9$ & $20.1 \pm 1.5$ & $55.2 \pm 1.4$ \\
$\mathrm{corr}(s_{\mathrm{inf}}, z)$ & $-0.064$ & $-0.202$ & $-0.464$ & $-0.807$ \\
$\mathrm{corr}(s_{\mathrm{unf}}, z)$ & $-0.051$ & $-0.057$ & $-0.120$ & $-0.235$ \\
Spread inf & $2.564 \pm .025$ & $2.544 \pm .033$ & $2.447 \pm .031$ & $2.290 \pm .020$ \\
Spread unf & $2.616 \pm .012$ & $2.619 \pm .028$ & $2.611 \pm .035$ & $2.586 \pm .016$ \\
\bottomrule
\end{tabular}
\end{table}

The value of the signal (VoS) rises monotonically from $1.6$, which is not distinguishable from zero, to $55.2$. The informed maker's directional quoting increases with its precision, and its spread tightens from $2.564$ to $2.290$. The uninformed spread stays essentially constant. Its per-seed slope on $p_1$ is $-0.066 \pm 0.050$ and equivalent to zero (TOST $p = 0.006$), whereas the informed maker's slope of $-0.613$ ($t = -10.4$) clearly fails the same test. The uninformed makers do respond to the rival through the direction of their quotes. Their skew tracks the latent state, which they can infer only from the rival's quotes, and this relation strengthens with the rival's precision: $\mathrm{corr}(s_{\mathrm{unf}}, z)$ falls from $-0.051$ to $-0.235$ across the sweep. Better information in the market therefore changes the direction of the uninformed quotes while their width stays the same.

Two further designs give the same picture. In the two-agent game the uninformed spread is $2.612 \pm 0.050$ at $p_1 = 0.8$ and $2.599 \pm 0.025$ at $p_1 = 1.0$, while the informed spread tightens by about five combined standard errors and the whole gain in payoff goes to the informed maker. With five seeds this cell lacks the power to pass the equivalence test (TOST $p = 0.105$), so we report it as consistent with a flat spread, without claiming equivalence. When the informed role is instead reassigned at random in each episode, with and without a public label identifying the informed maker, the uninformed spread passes the equivalence test in both cases (TOST $p = 0.0009$ and $p = 0.002$). The flat spread is therefore not a consequence of fixed roles.

\subsection{Mechanism}\label{sec:incidence}
A quoted spread reflects two considerations: the toxicity a maker expects on the fills it wins, and how strongly it competes for those fills. These correspond to the two forces of Section~\ref{sec:intro}, and a better-informed rival affects both. It leaves more of the toxic flow to the other makers, which favours wider quotes, and it quotes more tightly, so that a given width wins less flow, which favours tighter quotes. The uninformed spread changes only if one of these effects dominates. The rival cannot make an individual adverse fill more costly, because the price move after a fill is determined by the flow process, which no agent's information enters. Holding the toxicity of the flow fixed separates the informed-maker question from the informed-taker question: if a maker's signal could worsen the flow itself, the setting would again be one of informed takers.

\begin{figure*}[t]
\centering
\includegraphics[width=0.82\textwidth]{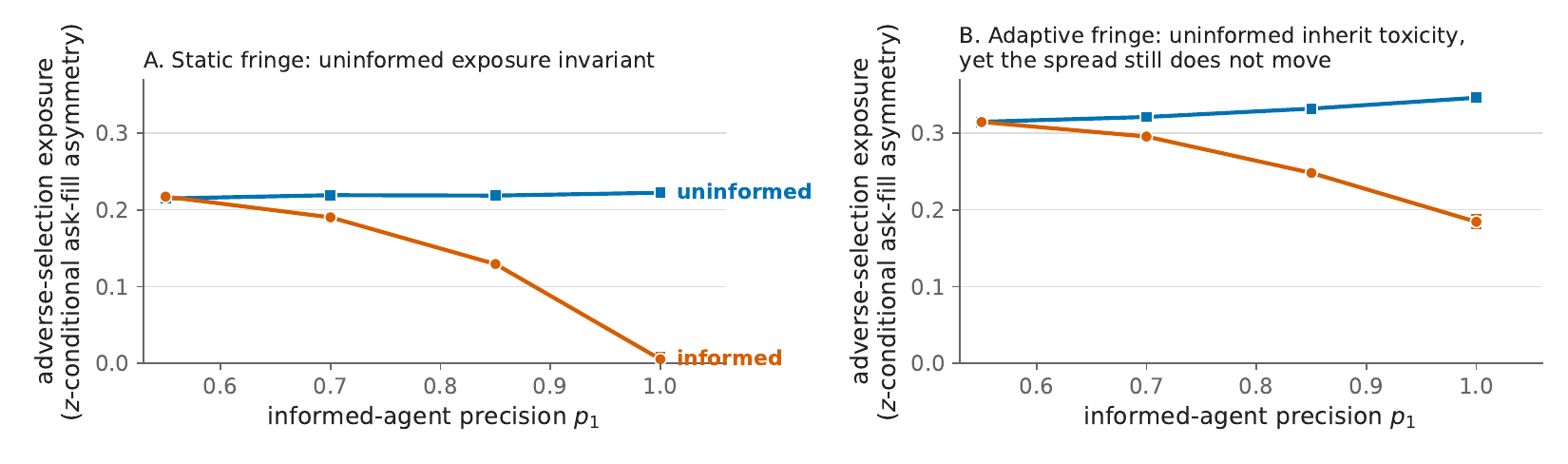}
\Description{Two line plots side by side. The horizontal axis is the informed maker's precision from 0.55 to 1.0; the vertical axis is adverse-selection exposure, the z-conditional ask-fill asymmetry, from 0 to about 0.35. Left panel, static fringe: the line for the uninformed makers stays flat at about 0.22, while the line for the informed maker falls from about 0.22 to about 0. Right panel, adaptive fringe: the uninformed line rises slightly from about 0.32 to about 0.35, while the informed line falls from about 0.32 to about 0.18.}
\caption{Adverse-selection exposure by role, measured as the $z$-conditional ask-fill asymmetry (fixed roles, ten seeds). \emph{Left, static fringe:} the informed maker's exposure falls to nearly zero, while the uninformed makers' exposure stays constant because the fringe absorbs the toxic flow that the informed maker avoids. The uninformed spread is flat over the same range (Table~\ref{tab:sweep}). \emph{Right, adaptive fringe ($\kappa = 1$):} the fringe reprices according to its own realised toxicity, quotes wide already at $p_1 = 0.55$ and absorbs less flow. Both kinds of maker therefore carry more toxicity at every precision, and in addition the uninformed makers' exposure rises with the rival's precision. Their spread still passes the equivalence test (Table~\ref{tab:af}), consistent with the rival tightening its quotes at the same time, which makes widening costly in lost flow.}
\label{fig:incidence}
\end{figure*}

We measure the protective force by a maker's \emph{exposure to toxic fills}, the fill asymmetry $\mathbb{E}[F^a \mid z{=}+1] - \mathbb{E}[F^a \mid z{=}-1]$. It is the excess of ask fills in the buy-pressure state over ask fills in the sell-pressure state, so a positive value means that the maker sells into rising markets. Figure~\ref{fig:incidence} plots this exposure by role under both fringe regimes; the between-seed SEM is at most $0.009$ throughout.

Under the static fringe the protective force is weak. The uninformed exposure changes only from $+0.215$ to $+0.222$, because the fringe absorbs the toxic flow that the informed maker avoids, while the informed maker's own exposure falls from $+0.217$ to $+0.005$. Most of the informed maker's gain of almost fifty per episode therefore comes from avoiding toxic fills, and only a small part from shifting them to its strategic competitors, who lose a few units in total. In this regime precision mainly determines the \emph{incidence} of adverse selection, that is, who bears it, and leaves the uninformed makers little additional toxicity to price. With informed takers the situation differs: a larger informed share raises the toxicity that quotes must price, and spreads widen \citep{glosten1985}; \citet{BarucciMathieuSB2025} report a slight widening and a clear fall in performance for an inventory-managing maker. Under the adaptive fringe of Section~\ref{sec:af} the fringe absorbs less, and the protective force becomes active: the uninformed exposure rises with the rival's precision from $+0.315$ to $+0.346$, a small change that is large relative to seed noise (paired $t = 9.1$ over ten seeds). Toxicity alone would now justify a wider quote, yet the uninformed spread does not increase. We interpret this as follows: the improvement in information that passes more toxic flow to the uninformed makers also leads the rival to quote more tightly, so that the competitive force offsets the protective one. The uninformed makers respond through instruments that do not cost order flow. They skew away from the side that the rival's quotes indicate as toxic, and they manage the resulting inventory.

Three further results are consistent with this interpretation. Among belief-conditioned deviations, defensive widening recovers about a quarter of the payoff deficit and never does better than directional imitation, which recovers about four fifths (Section~\ref{sec:deviations}). Pressure to widen does exist where these inexpensive responses are unavailable: the best state-independent constant quote, which can neither skew nor manage inventory, becomes wider as $p_1$ rises. Finally, Sections~\ref{sec:structure} and~\ref{sec:af} vary how strongly the makers compete for the marginal fill, and the response of the uninformed spread becomes negative where this competition is strongest.

\subsection{Deviations and Equilibrium Audits}\label{sec:deviations}
A learned profile need not be an equilibrium, and a flat learned spread does not imply that a flat spread is a best response. We therefore evaluate three families of deviations against the frozen learned competitors, the first two by paired simulation and the third by retraining, and then summarise the audits.

\emph{Constant spreads.} Replacing an uninformed agent with a constant symmetric quote $(\delta, \delta)$, $\delta \in [0.52, 2.0]$, does not improve on the learned policy at any precision; at each precision, the best constant loses 16 to 18 payoff units, about four standard errors of the paired difference. The best constant does become wider as $p_1$ rises, from $\delta \approx 1.58$ to the upper end of the grid. A quoter that cannot condition on any information therefore faces real pressure to widen, which the learned policies offset through state-dependent quoting at an unchanged average spread.

\emph{Belief-conditioned rules.} A deviator forms a belief $\hat z_t$ from the credibility-weighted mean of the competitors' skews and quotes $\delta^{b/a} = c \mp \theta \hat z_t + \omega |\hat z_t|$, so that $\theta = \omega = 0$ recovers the constant rule with per-side distance $c = 1.58$ (the best constant at $p_1 = 0.55$, tuned to a barely informed rival). No such rule improves on the learned policy; the best rule loses 1.5 to 6.5 units, within one paired standard error of zero. The decomposition of the rule is informative. At $p_1 = 1.0$ the constant rule falls $24.2$ units short of the learned policy---more than the $17.8$ lost by the best constant at that precision, since $c$ is no longer optimal there). The imitation term $\theta$, which skews in the direction implied by the rival's quotes, reduces this shortfall to $4.5$ units. The defensive-widening term $\omega$, which widens whenever the belief is strong in either direction, reduces it only to $18.2$ units. Following an informed rival's direction pays, in line with the price leadership in \citet{CalcagnoLovo2006}; defensive widening also helps, but much less, and it does not outperform the directional term at any precision. Best-response iteration over a constant symmetric spread for both uninformed seats likewise converges to spreads without a trend in precision (seed means $1.67$, $2.27$, $1.70$ and $1.87$, non-monotone and within iteration noise), so the restricted equilibrium of the model does not predict widening either.

\emph{Best responses.} Retraining a single uninformed maker in the fixed-role three-agent game against its frozen opponents, starting from its own weights, yields no gain in any cell: the mean paired regret per cell lies between $-6.2$ and $+4.7$ per episode at both precision extremes, and the retrained spreads are unchanged. The learned profiles are therefore local best responses. They are not global best responses. In a second test we freeze the informed maker and one uninformed maker and place in the remaining seat a deviator initialised from a \emph{blind} policy, that is, one trained without access to competitor quotes. After fine-tuning against the frozen opponents, this deviator earns $12$ to $25$ more per episode than the uninformed incumbent it replaces. It remains far below the informed maker, which earns about $94$ at $p_1 = 1.0$ against the deviator's $48$ to $60$, and its spread does not respond to $p_1$ either. In the reward decomposition, about $85$ percent of the deviator's advantage comes from holding roughly half the squared inventory and about $15$ percent from higher spread income. The gain appears to require opponents that do not adapt: neither joint training nor the warm-started audit reaches this profile, and a sighted entrant never beats the blind incumbent it replaces in the blind profile. The learned profile is therefore not a Nash equilibrium of the game. We interpret it as a \emph{learning equilibrium}, a stable point of the learning dynamics. The deviation changes the level at which these dynamics settle, while the comparative static is unaffected: the slope in $p_1$ is flat at the learned profile, at the deviator's profile and in the blind profiles.

\emph{Audit outcomes.} No retrained best response widens the spread systematically. Across the audited runs, retraining changes the spread by at most $\pm 6$ percent, and the most profitable exploit widens it by $5.4$ percent, which is within the seed noise of the runs it exploits. All reported statistics are unchanged when we restrict the sample to runs that pass the audit. The audit does reveal differences in the quality of learned play. 
In the two-agent game and in the role-randomised three-agent sweep, where we audit each maker of individual runs, regret is small at high precision ($-8$ to $-15$ percent at $p_1 = 1.0$, and $-2.4$ percent for the informed maker in the two-agent game) but mixed where the signal is weak: at $p_1 \in \{0.55, 0.7\}$, in one audited run in three the retrained maker earns 20 to 34 percent more than the policy it started from, against the same frozen opponents. Joint training, in which all makers learn at the same time, did not arrive at these profiles. Independent learning thus reaches basins of different quality where the signal is least informative, which is possible because IPPO does not guarantee global convergence. The signal-blind rollouts point in the same direction. Under role randomisation at $p_1 = 0.55$, the informed policy earns $5.04 \pm 1.33$ more per episode when its own signal is set to zero (one-sided $p = 7 \times 10^{-5}$), whereas at $p_1 = 1.0$ the same rollout loses $32.9$. The learner overreacts to a barely informative signal. We attribute this to the learning procedure and interpret the low-precision cells with this in mind.

\subsection{Market Structure and Capacity}\label{sec:structure}
\begin{figure*}[t]
\centering
\includegraphics[width=0.76\textwidth]{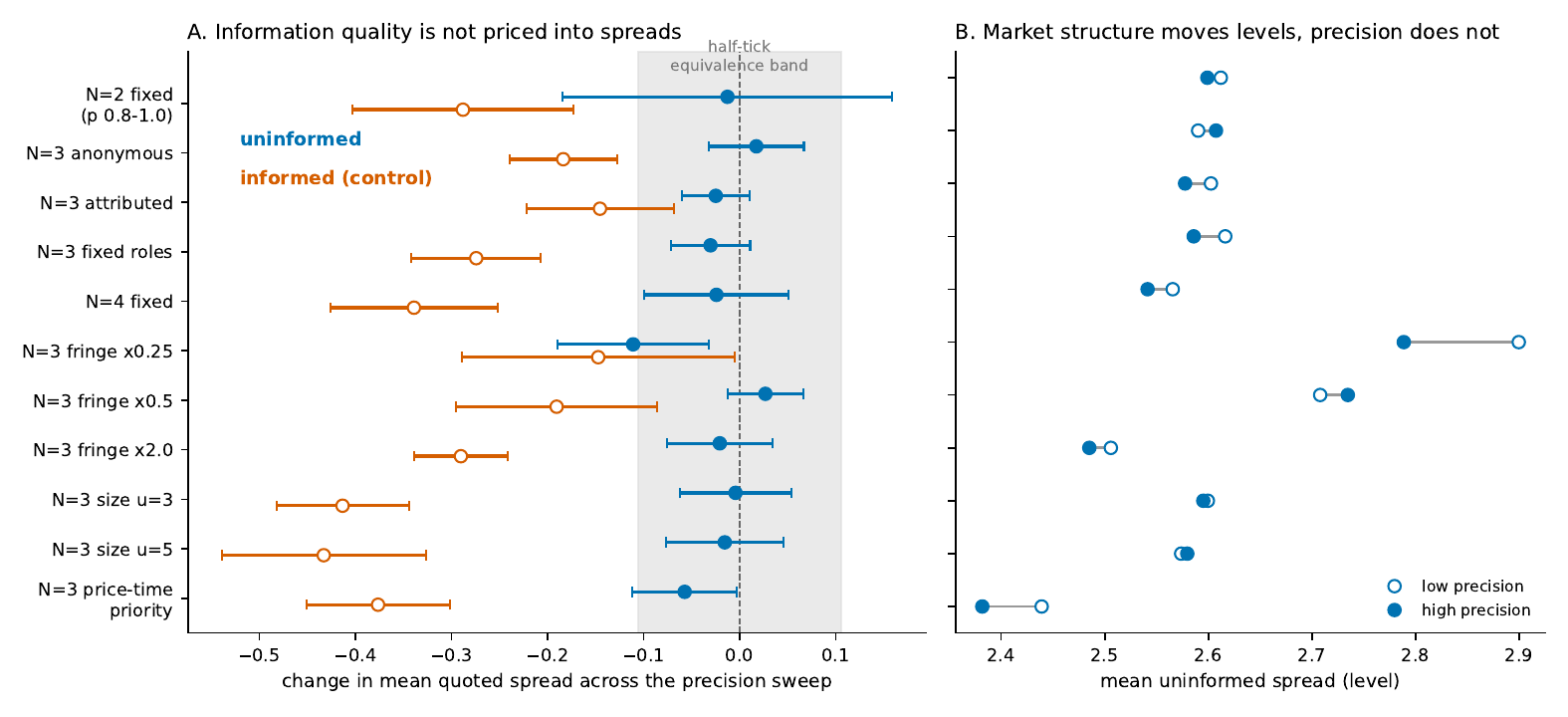}
\Description{Two-panel dot plot with one row per configuration: two makers with fixed roles, three makers with anonymous, attributed and fixed roles, four makers, fringe weights 0.25, 0.5 and 2.0, capacities 3 and 5, and price-time priority. Left panel: change in the mean quoted spread from low to high precision with 95 percent confidence intervals. The changes for the uninformed makers lie inside the shaded half-tick band around zero or below it and are never significantly positive; the changes for the informed maker are clearly negative in every row. Right panel: the uninformed spread level at low and high precision, between about 2.4 and 2.9, which varies more across configurations than between the two precisions.}
\caption{Spread response to the informed maker's precision for all market structures, capacities and execution rules in the paper (the adaptive-fringe sweep of Section~\ref{sec:af} is reported in Table~\ref{tab:af}). Panel A: per-seed change in the mean quoted spread from the low to the high end of each row's sweep (mean and 95 percent confidence interval; five seeds for $N{=}2$, eight for the $u{=}5$ low-precision cell, ten otherwise) for the uninformed makers (filled) and the informed maker (open, positive control), with the half-tick equivalence band. For rows swept over four precisions, the text reports slopes per unit of $p_1$, equal to these changes divided by the swept range of $0.45$. Panel B: level of the uninformed spread at both ends of the sweep. Across rows, the uninformed spread changes by at most about half a tick and never increases significantly, while the informed maker's spread tightens in every row.}
\label{fig:phase}
\end{figure*}

The fringe and the number of competitors could plausibly drive the result, so we vary both. We scale the fringe's execution weight by $0.25$, $0.5$ and $2.0$ relative to the baseline and add a configuration with $N{=}4$, each with fixed roles, ten seeds and $p_1 \in \{0.55, 1.0\}$. Figure~\ref{fig:phase} collects these together with all other configurations trained at both precision extremes, showing the response to precision in Panel A and the spread levels in Panel B.

These treatments change outcomes substantially, so the design is sensitive to market structure. Reducing the fringe weight from $\times 2.0$ to $\times 0.25$ raises the level of the uninformed spread from $2.506 \pm .021$ to $2.900 \pm .026$ and the uninformed payoff from $25.5$ to $86.4$; a fourth maker lowers the level to $2.566 \pm .021$; and the informed spread tightens with $p_1$ in every configuration. In none of them does the uninformed spread increase significantly with the rival's precision. Three configurations pass the equivalence test, namely fringe $\times 0.5$, fringe $\times 2.0$ and $N{=}4$ (TOST $p = 0.0007$, $0.0037$ and $0.019$). The only positive point estimate among them, $+0.027 \pm 0.017$ for fringe $\times 0.5$, lies within the margin and is not significantly different from zero. The thinnest fringe, $\times 0.25$, is the only configuration that fails the test, and it fails in the negative direction ($-0.111 \pm 0.035$, $t = -3.2$): its uninformed makers tighten along with the better-informed rival. Section~\ref{sec:af} examines this competitive regime in more detail.

Larger quote sizes give a similar result. Retraining with capacities $u = 3$ and $u = 5$ allows quantity effects that the one-unit cap rules out, and these effects appear: the uninformed toxic fill asymmetry rises from $+0.22$ to $+0.30$ and the value of the signal from $55$ to $73$. The uninformed spread again shows no response and passes the equivalence test at both capacities (TOST $p = 0.002$ and $0.005$), while the informed maker's tightening grows with capacity ($-0.27$, $-0.41$ and $-0.43$ for $u = 1$, $3$ and $5$). The $u{=}5$ cell at $p_1 = 0.55$ is evaluated on eight of its ten runs, and all differences involving it are paired within these eight. With all ten runs the uninformed spread still passes the equivalence test (TOST $p = 0.005$). Fills are far from saturated at every capacity. No fill occurs on either side in half of all steps, and the uninformed quote lies in the interior of the action grid.

\subsection{An Adaptive Fringe}\label{sec:af}
A further concern is that the baseline fringe is passive. It quotes at $1/k$ regardless of how toxic the flow becomes, although it is exactly the kind of break-even competitive segment that adverse-selection reasoning expects to widen. We therefore let the fringe respond. At each step the adaptive fringe widens above $1/k$ by $\kappa$ times its own recent average adverse markout per fill, where $\kappa = 0$ reproduces the static fringe exactly and $\kappa = 1$ gives the full response. The base margin is unchanged. If the flat spread depended on a fringe that cannot adjust, it should change in this treatment.

Table~\ref{tab:af} reports the sweep. At full strength the fringe's realised half-spread settles near $2.0$, compared with $1/k \approx 0.67$ in the static case, so the fringe no longer absorbs toxic flow at a fixed price. The spread and payoff levels of the strategic makers rise accordingly; the uninformed payoff reaches $86$ to $97$, against $39$ to $45$ under the static fringe. The per-seed slope of the uninformed spread on $p_1$, the same statistic as in Table~\ref{tab:sweep}, is $-0.033 \pm 0.084$ and passes the equivalence test (TOST $p = 0.026$). The informed maker's slope of $-0.41$ ($t = -4.9$) fails the same test, and the value of the signal rises from $-1$ to $+44$.

At half strength the competitive force becomes visible. The realised fringe half-spread is $0.67$, $1.20$ and $2.00$ for $\kappa = 0$, $0.5$ and $1$, so at $\kappa = 0.5$ the fringe quotes close to the strategic makers. The mean uninformed spread falls monotonically across precisions ($2.80$, $2.77$, $2.74$). The point estimate of the slope is negative ($-0.136 \pm 0.122$) and does not pass the equivalence test; some seeds tighten along with the better-informed rival and others do not. The clearest instance of this tightening remains fringe $\times 0.25$. Taken together, the treatments suggest that the regime depends on how strongly the fringe competes for the marginal fill, which its price and its weight determine jointly. When the fringe quotes far outside ($\kappa = 1$), the makers' rents are safe and the uninformed spread does not respond. When the fringe quotes at an intermediate distance ($\kappa = 0.5$) or carries little weight (fringe $\times 0.25$), the marginal fill is contested and the uninformed makers tighten. When the fringe is tight and at full weight, as in the static baseline, undercutting it attracts mostly toxic flow, and the uninformed spread again does not respond. In none of these regimes does the rival's precision widen the uninformed spread.

Two further results help locate the mechanism. First, a blind variant retrained at $\kappa = 0.5$, whose makers do not observe competitor quotes, shows no response to precision ($+0.02 \pm 0.07$, ten seeds) and quotes wider in level ($+0.12$ and $+0.30$, paired). The tightening at $\kappa = 0.5$ therefore depends on observing the rival's quotes; the fill experience alone, which the blind variant shares, does not produce it. Second, the repriced fringe measurably shifts toxicity onto the strategic makers ($t = 9.1$ for the rise in uninformed fill asymmetry, Section~\ref{sec:incidence}), yet the uninformed spread does not increase. The protective force is active in this treatment and is offset, which we attribute to the rival tightening at the same time. The flat spread is thus a property of the learned profiles of the strategic makers and does not depend on a fringe that cannot reprice.

\begin{table}[t]
\centering
\caption{Adaptive-fringe sweep (fixed roles, ten seeds per cell, $1{,}000$ evaluation episodes per seed; mean $\pm$ between-seed SEM; uninformed entries average the two uninformed makers). The $\kappa = 0.5$ arm was trained at three of the four sweep precisions. Last column: realised mean fringe half-spread, compared with the static value $1/k \approx 0.67$.}
\label{tab:af}
\footnotesize
\begin{tabular}{lccccc}
\toprule
$\kappa$ & $p_1$ & Spread unf & Spread inf & VoS & Fringe $\delta_{\mathrm{bg}}$ \\
\midrule
$1.0$ & $0.55$ & $2.894 \pm .021$ & $2.903$ & $-0.9$ & $1.96$ \\
$1.0$ & $0.70$ & $2.902 \pm .030$ & $2.884$ & $7.7$ & $1.96$ \\
$1.0$ & $0.85$ & $2.913 \pm .029$ & $2.847$ & $24.2$ & $1.97$ \\
$1.0$ & $1.00$ & $2.874 \pm .032$ & $2.712$ & $43.8$ & $2.00$ \\
$0.5$ & $0.55$ & $2.802 \pm .030$ & $2.767$ & $3.7$ & $1.20$ \\
$0.5$ & $0.85$ & $2.770 \pm .022$ & $2.614$ & $25.5$ & $1.22$ \\
$0.5$ & $1.00$ & $2.739 \pm .038$ & $2.510$ & $52.9$ & $1.24$ \\
\bottomrule
\end{tabular}
\end{table}

\subsection{Price-Time Priority}\label{sec:ptp}
A final concern is the fill rule. Under the contest allocation, a wider quote reduces exposure only in proportion to the flow it gives up, and this property alone might produce the flat spread. Under price-time priority, the execution rule of most lit equity and foreign-exchange venues, a maker that widens moves back in the queue and does not trade when a toxic sweep arrives, although it also gives up the harmless flow at the touch. Arguments for defensive widening, including the winner's curse of \citet{CalcagnoLovo2006}, rely on the best quote winning the trade, as in their first-price auction, and price-time priority shares this feature, so it is a natural place for the protective force to reappear.

We retrain the fixed-role game under an execution rule in which each buy (sell) order fills the tightest ask (bid) up to its capacity and then the next tightest, with ties broken at random as a proxy for time priority within the period; the fringe quotes at $1/k$. With this rule switched off, the environment is identical to the contest baseline. The per-seed slope of the uninformed spread on $p_1$ is $-0.113 \pm 0.043$ over ten seeds and passes the equivalence test (TOST $p = 0.016$). The point estimate is slightly negative, similar to the tightening in the contested regimes of Section~\ref{sec:af}. The informed maker's slope of $-0.81$ ($t = -10.2$) fails the same test, and the value of the signal rises to $+95$.

\section{Implications}\label{sec:practice}
The model is stylised, and its numbers should not be read as estimates for real markets. The comparisons above isolate mechanisms within the stated scope, in which the informed party is a competing liquidity provider. The implications below concern which quantities respond to a rival's information. They do not make claims about magnitudes and should be read as hypotheses for empirical work.

\emph{(1) Responses to a better-informed rival.} In our model, makers facing a better-informed rival protect themselves mainly through skew and inventory management. No tested deviation profits from widening. The blind deviator of Section~\ref{sec:deviations} gains mainly through inventory management, and its gain appears to require opponents that do not adapt. Constant quoters lose the mark-to-market income that the learned policies protect through state-dependent quoting. If this carries over to real markets, a desk whose competitor becomes better informed should see the effect mainly in its inventory profit and in the distribution of profits across makers, and less in its own markouts or quoted spreads.

\emph{(2) Market structure and rival information.} Varying the fringe's execution weight changes the uninformed payoff from $25$ to $86$ per episode, and making the fringe adaptive raises it from $39$--$45$ to $86$--$97$. Moving the rival's precision across its full range changes the uninformed payoff by about six units (Table~\ref{tab:sweep}). In our model, the structure of the venue matters considerably more for the uninformed makers' profits than the information quality of their competitor.

\emph{(3) Competitor-quote features in learned quoting policies.} In Section~\ref{sec:deviations} a deviator initialised from a blind policy earns substantially more than the incumbent it replaces, mostly through lower inventory risk at nearly unchanged spread income. For a desk with a learned quoting engine, this suggests measuring how much inventory risk the engine takes on in response to competitor-quote features and how much these features contribute to profit. In our model the added risk outweighs the contribution.

\emph{(4) Detecting maker-side information asymmetry.} In our model, moving the informed maker's precision from uninformative to perfect changes the uninformed spread by at most about half a tick and never increases it significantly, so the rivals' quoted spreads reveal little about the informed maker's advantage. This differs from models with informed takers, in which a growing informed share widens quoted spreads \citep{glosten1985}. The informed maker's own quotes do respond: at $p_1 = 1$ its spread is $2.29$, against $2.59$ for the uninformed makers, and $\mathrm{corr}(s, z)$ moves from $-0.06$ to $-0.81$ across the sweep. This is an imperfect indicator, because tight directional quoting has other causes and the uninformed makers partly imitate the direction. Payoffs respond more clearly, rising from about the uninformed level to more than twice that level. Venues and regulators interested in informationally advantaged liquidity providers could therefore monitor the distribution of maker profits and fill quality. Displayed capacity is exogenous in our model, so we cannot assess quoted depth as an indicator.
\balance

\section{Conclusion}\label{sec:conclusion}
In the class of markets studied here, the spreads of uninformed makers respond to competition and inventory risk but not to the information quality of a rival maker. We support this with equivalence tests, a positive control in every sweep, and deviation audits, none of which finds a profitable widening. Our measurements are consistent with a competitive trade-off. A better-informed competitor mainly avoids toxic flow, and where it passes more toxic flow to the others, it also quotes more tightly, so that additional width would cost the uninformed makers more order flow than it saves. Skew and inventory management are the less costly responses, and the rival's information changes the distribution of profits while quoted spreads remain nearly unchanged. The result holds with a fringe that reprices after toxic fills and under price-time priority. Our conclusions are limited by the stylised flow process, the contest allocation used in most treatments, and the fact that the learned profiles are not Nash equilibria. The deviation analysis suggests that reacting to competitor quotes adds inventory risk that the information in those quotes does not compensate, so a regulariser on the reactivity to competitor-quote features is a natural next step. Further extensions are a market with several informed makers and one with informed takers alongside informed makers, in which the informed share of the flow varies as in \citet{BarucciMathieuSB2025}, so that the amount and the incidence of toxicity change together.

\section*{CRediT Author Statement}
\textbf{Oliver James Oberle:} Conceptualization, Methodology, Software, Validation, Investigation, Writing--Original Draft, Visualization. \textbf{Damien Berriaud:} Conceptualization, Methodology, Validation, Writing--Review \& Editing, Supervision. \textbf{Roger Wattenhofer:} Supervision, Funding Acquisition.

\bibliographystyle{ACM-Reference-Format}
\bibliography{references_camera_ready}

\end{document}